\documentclass[12pt]{iopart}

\ifx\pdftexversion\undefined
  \usepackage[dvips]{graphicx}
\else
  \usepackage[pdftex]{graphicx}
\fi
\usepackage{hyperref}
\usepackage{floatflt}
\newcommand{\ee}[1]{\!\times\!10^{#1}}

\begin{document}
\title[An MCMC search for GWs of uncertain frequency from a targeted neutron
star]{A time-domain MCMC search and upper limit technique for gravitational waves of uncertain
frequency from a targeted neutron star.}
\author{John Veitch$^1$, Richard Umst\"atter$^3$, Renate Meyer$^3$, Nelson
Christensen$^2$, Graham Woan$^1$}
\address{$^1$Department of Physics and Astronomy, Kelvin Building, University of
Glasgow, Glasgow G12 8QQ, UK \\
$^2$Physics and Astronomy, Carleton College, Northfield, MN 55057, USA\\
$^3$Department of Statistics, University of Auckland, Auckland, New Zealand}
\ead{jveitch@astro.gla.ac.uk}

\begin{abstract} It is computationally expensive to search
the large parameter space associated with a gravitational wave
signal of uncertain frequency, such as might be expected from the
possible pulsar generated by SN1987A. To address this difficulty
we have developed a Markov Chain Monte Carlo method that performs
a time-domain Bayesian search for a signal over a 4\,Hz frequency
band and a spindown of magnitude of up to $1\ee{-9}$\,Hz/s. We use Monte Carlo
simulations to set upper limits on signal amplitude with this technique, which we intend to apply to a gravitational wave search.
\end{abstract}


\section{Introduction}
Recent targeted searches for gravitational waves from radio
pulsars have needed to consider only four signal parameters:
$h_0$, the signal amplitude, $\cos{\iota}$, the inclination angle
of the pulsar toward the line of sight, $\psi$, the polarisation
angle of the gravitational wave and $\phi_0$, the initial phase of
the signal.  In the absence of a detection, it is possible to set
an upper limit on $h_0$, and therefore on the ellipticity of the
pulsar in question.  This is accomplished by defining a Bayesian
posterior probability density function (PDF) of the four signal
parameters and marginalising it numerically on a grid. In the
time-domain, this search technique relies on precise heterodyning
of data from the gravitational wave detector, which itself
requires accurate information on the phase evolution of the
signal.  For known pulsars this is easily obtained by monitoring
the timing of their radio wave pulses, which directly provides the
frequency and frequency derivatives of the gravitational signal
(emitted at twice the object's rotation frequency) \cite{S1TDS}.
However, there are objects whose frequencies are not known
accurately, such as the reported remnant of Supernova 1987A
\cite{Middleditch}. To perform a search over a range of
frequencies would require a grid on frequency and frequency
derivative with resolutions approximately $1/T_{\rm obs}$ and
$1/T_{\rm obs}^2$ respectively, which makes the number of search
templates very large ($\sim4.5\ee{11}$ for LIGO S3 data). We
present an adaptation of this search based on a Markov Chain Monte
Carlo (MCMC) method which does not require an exhaustive
examination of the parameter space and therefore is able to search
a range of frequencies in a reasonable time.

\section{MCMC parameter estimation}
After heterodyning at close to the expected signal frequency, the
gravitational wave signal from a rotating triaxial neutron star
has the form
\begin{equation}
y(t_k;{\bf a})=\Big(\frac{1}{4}F_+(t_k;\psi)(1+\cos^2\iota)-\frac{i}{2}
F_\times(t_k;\psi)\cos{\iota}\Big)h_0e^{i\Phi(t_k)}
\end{equation}
where $F_+$ and $F_\times$ are the beam patterns for the $+$ and $\times$
polarisations, and $\Phi(t)$ is the phase of the signal given to the first two
Taylor expansion terms by
\begin{equation}
\Phi(t)=\phi_0+2\pi[\delta{}f(T-T_0)+\frac{1}{2}\delta{}\dot{f}(T-T_0)^2].
\end{equation}
Here, $\delta{}f$ is the deviation of the signal frequency from
the heterodyne frequency, and $\delta\dot{f}$ is the deviation
from first derivative of the heterodyne frequency \cite{JKS}. By
including these two parameters we can search around the heterodyne
frequency for a signal. The heterodyned data is reduced to one
complex sample per minute, $B_k$, with variance $\sigma_k$,
allowing a frequency range of $[-1/120,1/120]$\,Hz around the
heterodyne frequency to be searched. The search in $\delta\dot{f}$
is limited to $[-1\ee{-9},1\ee{-9}]$\,Hz/s which is likely to
include any possible pulsar spindown rates. The search proceeds by
heterodyning 480 separate frequency bands at intervals of
$1/120$\,Hz, and running parallel searches on each band. In this
way we complete a search over a 4\,Hz range -- suitable for the
putative SN1987A pulsar.

The probability of a particular combination of the six parameters
${\bf a}$ representing a signal in the data $\{B_k\}$ is
\begin{equation}
p({\bf a}|\{B_k\})\propto p({\bf a})\exp{}
\left[-\sum_k\frac{|B_k-y(t_k;{\bf a})|^2}{2\sigma_k^2}\right],
\end{equation}
which is the prior probability of ${\bf a}$ multiplied by its
likelihood. This proportional definition is adequate for our
application, as we will only be evaluating ratios of
probabilities.

We now have a joint posterior probability distribution over the
six parameters in vector ${\bf a}$, but in the presence of a
signal the probability will be strongly concentrated around the
signal parameters. The Markov Chain Monte Carlo method takes
advantage of this fact by preferentially sampling the areas of
greatest probability density, in proportion to that probability
\cite{MCMCbook,6PCQG}. In this way we can approximate the
posterior distribution $p({\bf a}|\{B_k\})$ by allowing the Markov
Chain to accumulate many samples, where the density of samples in
an area is then proportional to the probability density in that
area. In order to accomplish this, a Markov Chain in state ${\bf
a}$ chooses a candidate sample ${\bf a'}$ from proposal
distribution $q_1({\bf a'}|{\bf a})$, and then accepts the
candidate as the next state of the chain with probability
\begin{equation}
\alpha_1({\bf a'}|{\bf a})={\rm min}\left[1,\frac{p({\bf a'})
p(\{B_k\}|{\bf a'})q_1({\bf a}|{\bf a'})}{p({\bf a})
p(\{B_k\}|{\bf a})q_1({\bf a'}|{\bf a})}\right].
\end{equation}
If the proposal is not accepted, the current state ${\bf a}$ is
recorded as a sample again and the process repeats. The proposal
distribution $q_1({\bf a'}|{\bf a})$ is a multivariate normal
distribution, with covariances determined by the correlation
between parameters in trial runs. In practice, the highly
correlated parameters $h_0$ and $\cos\iota $ are replaced with a
re-parameterisation, such that $a_1=\frac{1}{4}h_0(1+\cos^2\iota)$
and $a_2=\frac{1}{2}h_0\cos\iota $. The frequency parameters are
also changed to $f_{\rm
start}=\delta{}f+\frac{1}{2}\delta\dot{f}\times{}t_{\rm start}$
and $f_{\rm end}=\delta{}f+\frac{1}{2}\delta\dot{f}\times{}t_{\rm
end}$, being the signal frequency at the start and end times of
the data.

It is important that the chain explore the parameter space
adequately, in order to find the area of high probability. To this
end, we have included a delayed rejection stage of the algorithm,
so that if a proposal ${\bf a'}$ is rejected, a new candidate is
generated from distribution $q_2({\bf a''}|{\bf a'},{\bf a})$,
which has a narrower width and makes more conservative steps.
These are accepted with probability
\begin{equation}
\fl \alpha_2({\bf a''}|{\bf a})={\rm min}\left[1,\frac{p({\bf
a''})p(\{B_k\}|{\bf a''})q_1({\bf a'}|{\bf a''})q_2({\bf a}|{\bf
a'},{\bf a''})(1-\alpha_1({\bf a'}|{\bf a''}))}{p({\bf
a})p(\{B_k\}|{\bf a})q_1({\bf a'}|{\bf a})q_2({\bf a''}|{\bf
a},{\bf a'})(1-\alpha_1({\bf a'}|{\bf a}))}\right],
\end{equation}
where the proposal distributions are included to satisfy the
principle of detailed balance. This allows the chain to make small
steps where a large step would be rejected.

In addition to these measures, there is also a burn-in period at
the start of each MCMC run where the exponent in Eqn.~3 is
multiplied by an inverse temperature factor $\beta$ in the
proposed step. Initially $\beta=0.01$, and gradually increases
during the burn-in period to $\beta=1$, where Eqn.~3 is restored.
This decreases the likelihood of the chain getting stuck in a
local maximum of probability without exploring the space
adequately. No samples from the burn-in period are used in
calculating the final PDF, as they do not represent the target
distribution $p({\bf a}|\{B_k\})$. In our implementation the
burn-in lasts for $1\ee{6}$ iterations, followed by 100\,000
iterations with every 50th used as a sample, so as to reduce
correlation between samples.  The program is implemented in C,
using the LIGO Algorithm Library to calculate the beam patterns
and time delays necessary for analysing a real signal \cite{LAL}.

\section{Setting upper limits}
In the standard time-domain search, if a signal is not detected it
is useful to set an upper limit on gravitational radiation emitted
by the source under examination. This translates as an upper limit
on the ellipticity of the pulsar which can be used to constrain
physical models of a neutron star. This is achieved by
marginalising the posterior PDF over $\cos\iota $, $\psi$ and
$\phi_0$, leaving a distribution on $h_0$ which can then be
integrated upward from $h_0=0$ until 95\% of the probability is
included in the interval. The upper limit of the interval is then
the 95\% upper limit on $h_0$ for the target.

In the MCMC routine however, if the signal is below a certain
threshold then the chain may not converge on the correct mode of
the signal in the PDF \cite{gilks96}. This is partly due to the
narrowness of the mode in $\delta{}f$ and $\delta\dot{f}$, where
the bulk of the probability lies within a single frequency bin of
width $1/T_{\rm obs}$ with a very small attraction area.

\begin{figure}[!htbp]
\includegraphics[width=0.5\textwidth,height=5.5cm]{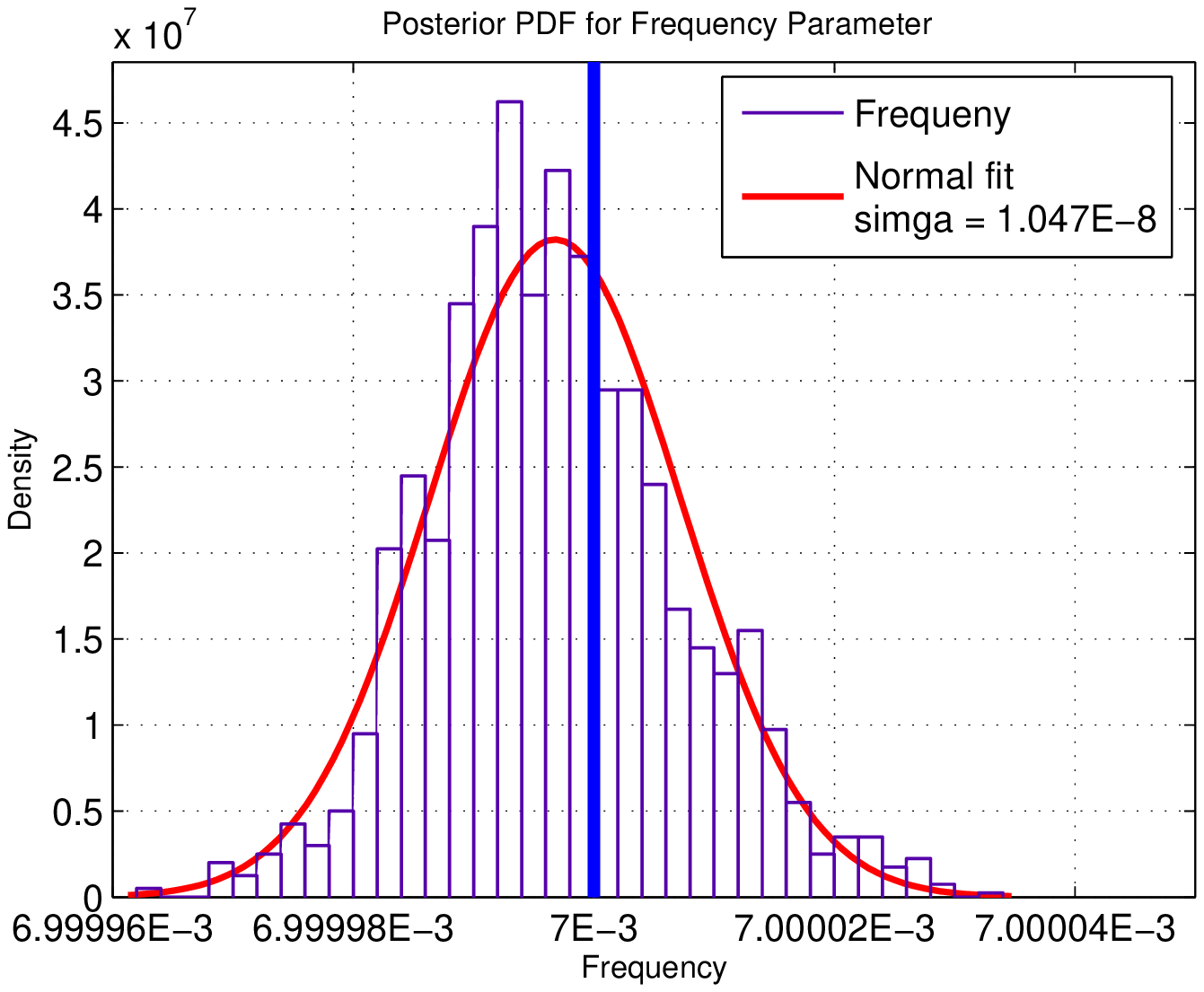}
\includegraphics[width=0.5\textwidth,height=5.5cm]{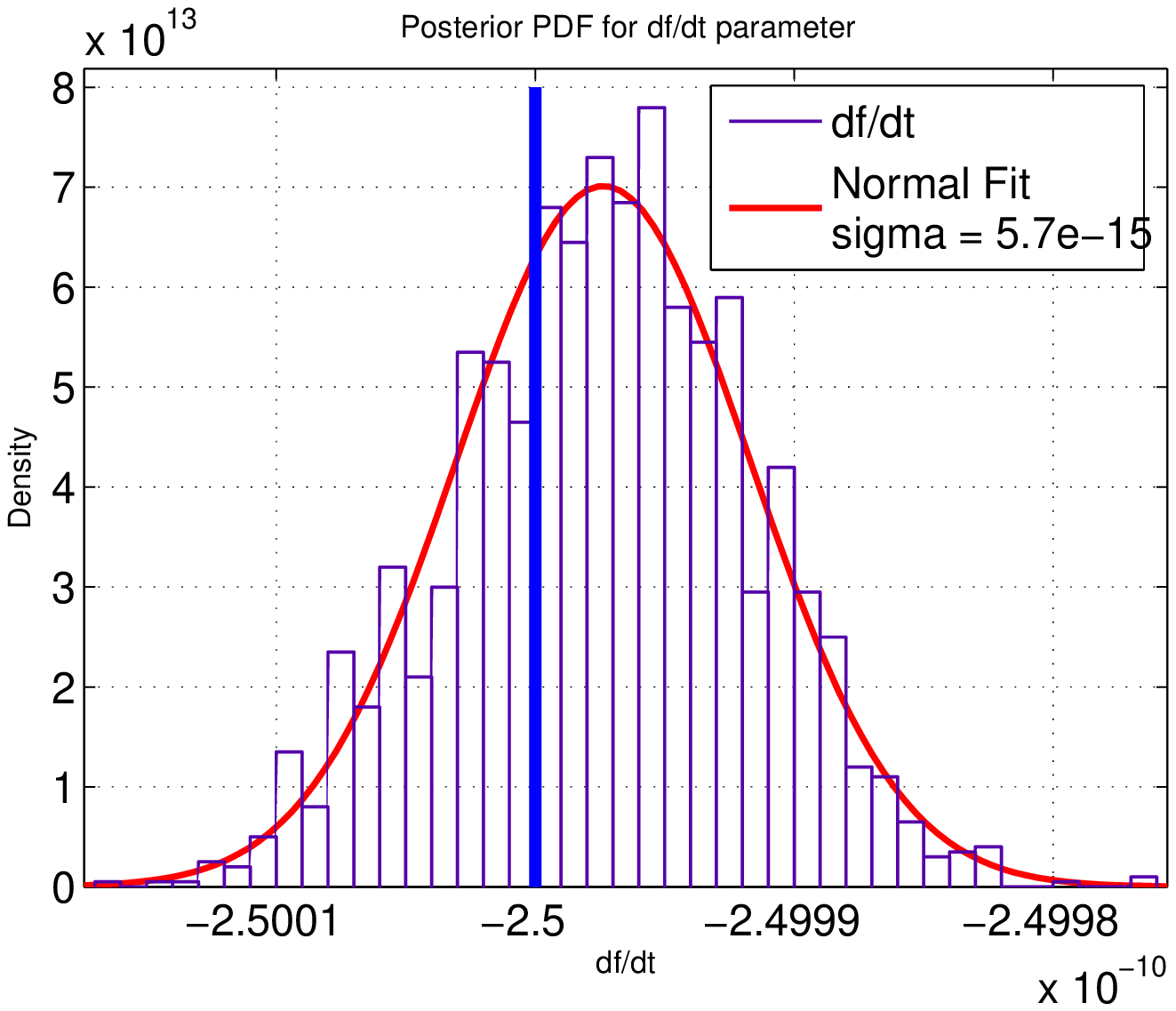}
\begin{center}
\caption{The posterior PDFs (bars) for a 2000 sample chain which
has converged on the injected values of $\delta{}f=7\ee{-3}$\,Hz
and $\delta\dot{f}=-2.5\ee{-10}$\,Hz/s (vertical lines), and the
best normal fits, having standard deviations $1.047\ee{-8}$\,Hz
and $5.7031\ee{-15}$\,Hz/s respectively.}
\end{center}
\end{figure}

A weak signal will raise probability only slightly in this bin, so
that even if the chain were to find the mode, it has a probability
of jumping out and not returning. These factors therefore preclude
the evaluation of an upper limit simply by marginalising the
posterior PDF when there is no detection. Instead, we performed a
Monte Carlo simulation by injecting signals of known parameters
into noise, and running the MCMC code to try and recover them.
Signals are injected with known parameters by calculating the
value of the signal at each timestamp $t_k$ using LAL and adding
it to each sample $B_k$ in the input data file. Real data would
have to be heterodyned prior to this step, but the artificial
noise is generated at $1/60$\,Hz so heterodyning is unnecessary in
the simulations. It can be shown that the probability of detecting
a signal using our method depends strongly on the injected values
of $h_0$ and $\cos\iota $, and much more weakly on the value of
$\psi$. To determine our upper limit for a particular set of data,
we inject signals of varying $h_0$  and $\cos\iota $ into the
noise $\sigma_k$, then analyse the results of the MCMC routine to
determine if it has detected the signal. This is accomplished by
observing the output PDFs in $\delta{}f$ and $\delta\dot{f}$, and
determining if the chain has converged or not. If the injected
value lies within three standard deviations of the mean, and the
standard deviations themselves are less than one frequency bin in
size, the chain is judged to have converged. If the chain has not
converged, the samples are distributed widely over the entire
$\delta{}f$ range, and the standard deviation is typically five
orders of magnitude higher. Fig.~1 shows the posterior PDFs in the
$\delta{}f$ and $\delta\dot{f}$ parameters for a chain converged
on a signal with injected parameters $h_0=1.3$, $\cos\iota =0$,
$\psi=0.281$, $\phi_0=4.234$, $\delta{}f=7\ee{-3}$\,Hz and
$\delta\dot{f}=-2.5\ee{-10}$ \,Hz/s, where the noise level was
$\sigma=1.0$. Fig.~2 shows the same type of plot for a signal with
$h_0=0.7$ and otherwise identical parameters; this chain failed to
converge, so there is no concentrated mode in the PDF. The
attempted fit to a normal distribution to test convergence is
therefore very wide, and poorly fitted.

\begin{figure}[!htbp]
\includegraphics[width=0.5\textwidth,height=5.5cm]{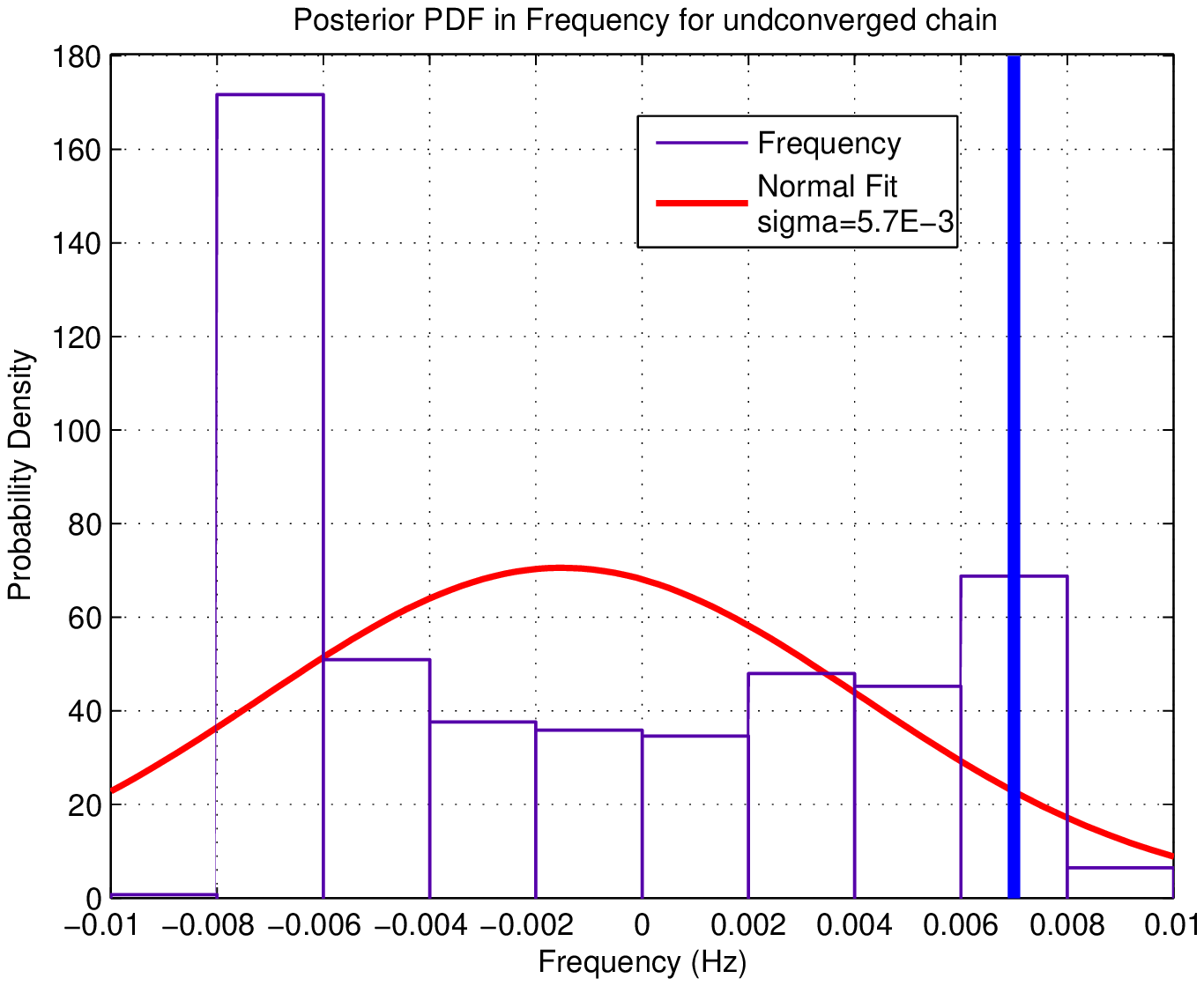}
\includegraphics[width=0.5\textwidth,height=5.5cm]{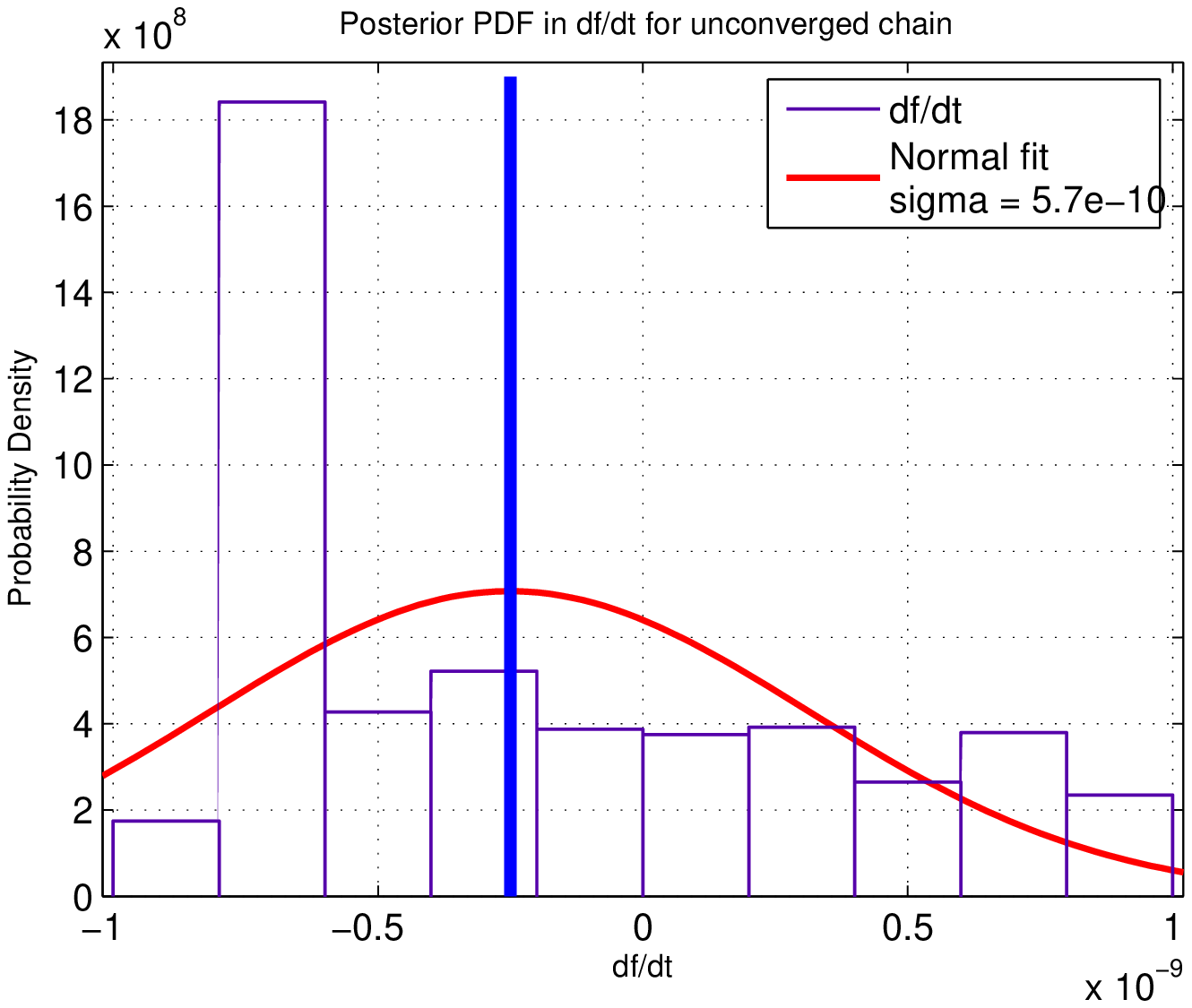}
\begin{center}
\caption{The $\delta{}f$ and $\delta\dot{f}$ posterior PDFs (bars)
for a 2\,000 sample chain which failed to converge on the injected
parameters (vertical lines). This distribution is clearly
distinguishable from the converged case using the criteria defined
above.}
\end{center}
\end{figure}

The injections are repeated, using a different starting point in
parameter space for the MCMC chain each time, and an empirical
probability of signal recovery $P(\rm detection|h_0,\cos\iota
,\sigma_k)$ for each point on the ($h_0$, $\cos\iota $) plane is
obtained from the set of results. Fig.~3 shows the result on the
($h_0$, $\cos\iota $) plane of performing this procedure on
64\,000 samples of white noise with a variance of $\sigma=1.0$.
The dark region shows where no signal is detectable, above which
is a small transition zone where there is a finite probability of
detection. With increasing iterations of the chain, the width of
this zone reduces - we are using a chain of 1\,100\,000
iterations, of which the first 1\,000\,000 are burn-in time and
the last 100\,000 used for sampling the PDF. The white area above
this transition zone represents signals that are detected with
very high probability. These are strong enough to cause the chain
to converge on them in every instance of the Monte Carlo
simulation. It is clear from the figure that this is strongly
dependent on $\cos\iota $, as this acts a weighting factor between
the power in the real and imaginary parts of the heterodyned
signal. The distribution is symmetric about $\cos\iota =0$, so
only the positive half was calculated.

\begin{figure}[!htb]
\begin{center}
\includegraphics[width=0.5\textwidth]{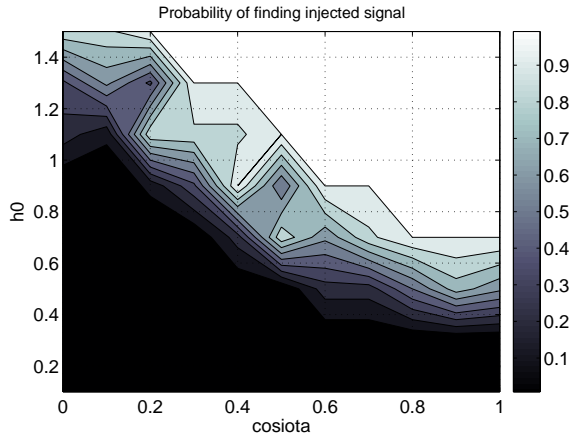}
\caption{Results of Monte Carlo evaluation of probability of
detection of a signal injected with varying values of $h_0$ and
$\cos\iota $. This shows the strong dependence on both these
parameters. The transition from detectable to undetectable occurs
over a narrow region for a particular value of $\cos\iota $, but
decreases in $h_0$ as $\cos\iota $ tends to 1.}
\end{center}
\end{figure}

This result can then be marginalised by summing over $\cos\iota $
and normalising to give a distribution on $h_0$, as required for
an upper limit. The probability of there being an undetected
gravitational wave is then $1-P(\rm detection|h_0,\sigma_k)$, and
the upper limit $h_{95\%}$ satisfies $1-P(\rm
detection|h_{95\%},\sigma_k)=0.95$. Note that $P(\cdot|\cdot)$ is
a probability rather than a probability density. The upper limit
determined from this stage is applicable to each of the 480 bands
in the 4\,Hz frequency interval because the values of $\sigma_k$
are determined by estimating the noise over the entire 4\,Hz band
at the heterodyne stage. The heterodyne then performs filtering of
this band to calculate the signal $B_k$ in each $1/60$\,Hz
interval, from which the 480 bands are derived.

\begin{figure}[!htb]
\begin{center}
\includegraphics[width=8cm]{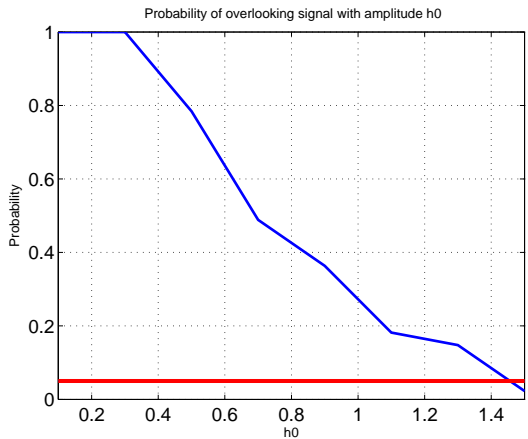}
\caption{The marginalisation of Fig.~3 onto the $h_0$ axis. This shows that
the 95\% confidence (represented by the horizontal line) may be drawn at $h_{95\%}=1.44$.}
\end{center}
\end{figure}

Fig.~4 shows this final distribution on $h_0$, where the
horizontal line represents 95\% confidence in detection, required
for a 95\% upper limit, which is achieved at $h_0=1.44$. It is
useful to express this in terms of noise power spectral density,
$S_h$, yielding the relation
\begin{equation}
h_{95\%}=515.2\sqrt{\frac{S_h}{T_{\rm obs}}} {Hz}^{-\frac{1}{2}}.
\end{equation}
where the 512.2 is an indicator of the sensitivity of the search,
empirically derived from the simulation.

For a realistic example, consider a noise power spectral density
of $S(f)=6.6\ee{-44}\,{\rm Hz}^{-1}$, or $h(f)=2.57\ee{-22}$, and a
data run of 44 days, the upper limit would then be approximately
\begin{equation}
h_{95\%}=515.2\sqrt{\frac{6.6\ee{-44}}{3801600}}=6.8\ee{-23} {Hz}^{-\frac{1}{2}}.
\end{equation}

\section{Summary}
This MCMC search method is capable of reliably detecting a signal
above a certain threshold and estimating its parameters including
frequency and spindown. The computing time it takes to perform the
search is dependent only on the amount of data and the number of
MCMC iterations desired, which provides significant improvement
over the previous time-domain targeted search\cite{S2TDS}.  With
the current version of the pipeline, to analyse the entire 4 Hz
window would take about 17\,000 CPU hours on a 1.8\,GHz Athlon
processor, which is approximately two days on the 366-CPU Merlin
cluster at AEI Golm. In the absence of a signal an upper limit can
be set using Monte Carlo injections, which require a similar
amount of processing on top of the search itself. Additional work
is under way to tune the algorithm further, which may lead to
improvements in sensitivity and speed of execution.

We hope that this technique could be applied to a search for
gravitational radiation from possible pulsar in the remnant of
SN1987A at $935\pm2$\,Hz, and in the absence of detection place
upper limits on the gravitational radiation emitted by this object
if it were triaxial. The method could also be easily applied to
other similar objects.

\verb''\ack
This work was supported by National Science Foundation grant
PHY-0244357, the Royal Society of New Zealand
Marsden fund award UOA204, Universities UK, PPARC and the University of
Glasgow. We would like to also thank Peter Saulson for reading and 
commenting
on the manuscript.


\section*{References}
\begin{thebibliography}{99}

\bibitem{S1TDS}
Abbot B \etal (The LIGO Scientific Collaboration) 2004 {\it Phys. Rev. D} {\bf
69} 082004

\bibitem{6PCQG}
Richard Umst\"atter, Renate Meyer, R\'ejean J Dupuis, John Veitch, Graham Woan and
Nelson Christensen 2004
{\it Class. Quantum Grav.} {\bf 21} No 20 S1655-S1665

\bibitem{Middleditch}
J. Middleditch, J. A. Kristian, W. E. Kunkel \etal. Rapid photometry of
supernova 1987A: a 2.14 ms pulsar? {\it New Astronomy} {\bf 5}:243-283, August
2000

\bibitem{MCMCbook}
D. Gamerman. {\it Markov Chain Monte Carlo: Stochastic Simulation for Bayesian
Inference}. Chapman \& Hall, 1997

\bibitem{LAL}
\url{http://www.lsc-group.phys.uwm.edu/lal/}

\bibitem{gilks96}  W.R. Gilks and S. Richardson and D.J. Spiegelhalter, {\it
Markov Chain Monte Carlo in Practice} (Chapman and Hall, London, 1996).

\bibitem{JKS} Jaranowski P, Kr\'olak A and Schutz B 1998 {\it Phys. Rev. D} {\bf
58} 063001

\bibitem{S2TDS}
Abbott B \etal (The LIGO Scientific Collaboration), Kramer M and
Lyne A, accepted by {\it Phys. Rev. Lett.} {\it Preprint}
\href{http://uk.arxiv.org/abs/gr-qc/0410007} {gr-qc/0410007}.
\end{thebibliography}
\end{document}